\documentstyle[preprint,aps]{revtex}
\begin{document}
\title{\bf Repulsion of Single-well Fundamental Edge Magnetoplasmons\\ In
Double Quantum Wells }
\author{O. G. Balev$^{*}$ and P. Vasilopoulos$^\dagger$} %$\dag$}
\address{$^{*}$Institute of Semiconductor Physics, National
Academy of Sciences,\\
45 Pr. Nauky, Kiev 252650, Ukraine\\
\ \\
$\ ^\dagger$Concordia University, Department of Physics, \\
1455 de
Maisonneuve Blvd O, Montr\'{e}al, Qu\'{e}bec, Canada, H3G 1M8}
%\end{center}
%\end{instit}
\date{July 31, 1998}
\maketitle
\begin{abstract}
A {\it microscopic} treatment of fundamental edge magnetoplasmons (EMPs)
along the edge of
a double quantum well (DQW) is presented for strong
magnetic fields, low temperatures, and  total filling factor $\nu=2$. 
It is valid for lateral confining potentials 
that Landau level (LL) flattening can be neglected. 
The cyclotron and Zeeman  energies are assumed larger than the DQW
energy splitting $\sqrt{\Delta^2 +4T^2}$, where $\Delta$ is the   splitting
of the isolated wells and $T$   the tunneling matrix element.
%hen calculated unperturbed density profile is sharp at the edge.
Using a random-phase approximation (RPA), which includes
local and nonlocal contributions to the current density,
it is shown that for negligible tunnel coupling $2T \ll \Delta$ the
inter-well Coulomb coupling leads to two DQW fundamental
EMPs which are strongly renormalized in comparison with
the  decoupled,  single-well  fundamental EMP.
These DQW modes can be modified further upon varying
the inter-well distance $d$,  along the z axis, and/or the
separation of the wells' edges $\Delta y$ along the $y$ axis.
The charge profile of the {\it fast} and {\it slow} DQW mode varies,
respectively, in an {\it acoustic} and {\it optical} manner along
the $y$ axis and is not smooth on the  $\ell_{0}$ scale.
For strong tunneling $\Delta\alt 2T$ these DQW modes
are essentially modified when $\Delta$ is changed by applying
a transverse electric field to the DQW.
\ \\
PACS\ \ 73.20.Dx, 73.40.Hm
\end{abstract}
%\end{document}
%\clearpage
\section{INTRODUCTION.}

Recently  the interest in %edge magnetoplasmons (
EMPs has been increased considerably \cite{1}-\cite{11}; for a discussion
of older studies see, e.g., Ref. \cite{1}. 
Initially EMPs were theoretically considered  only within
essentially classical models \cite{1},\cite{7} 
%which imply a specific edge-wave mechanism,
in which  the edge does not vary but the charge density
profile at the edge does.
Except for the complications due to the long-range Coulomb
interaction, this edge-wave mechanism is the magnetic
analog  of  that for the Kelvin wave \cite{11a}, i.e.,  
the chiral wave  that occurs in  tides at the edge of  a rotating "shallow"  
sea, the angular velocity of the sea being the analogue of the
cyclotron frequency.
More recently a distinctly different, fully quantum mechanical
edge-wave mechanism for EMPs was provided by the models of Refs.
\cite{3a}-\cite{3c} and \cite{7a},\cite{7b}. In these models typically 
only the edge position of an incompressible two-dimensional electron gas 
(2DEG) varies;  with respect to that the density profile is that of the
undisturbed 2DEG \cite{10a}. 

In Refs. \cite{10a}-\cite{11} we proposed
a quasi-microscopic model that effectively incorporates both  
 edge-wave mechanisms mentioned above. 
%In  Ref. \cite{11}  we have 
In this model we treated EMPs 
$\propto A(\omega,q_{x},y) \exp[-i(\omega t-q_{x} x)]$,  for a single
quantum well  with $\nu=1$ and very
low temperatures $T$,  for which the unperturbed density profile
drops sharply at the edges on a length of the order of
magnetic length $\ell_{0}$.
Such a profile is valid for $k_{B}T\ll \hbar v_{g}/\ell_{0}$, where
$v_{g}$ is the group velocity of the edge states,
and the unperturbed electron density $n_{0}(y)$, normalized to the
bulk value $n_{0}$, is calculated as
$n_{0}(y)/n_{0}=\{1+\Phi[(y_{re}-y)/\ell_{0}]\}/2$, where $y_{re}$ is
the coordinate of the right edge and $\Phi(y)$ the probability integral.
However, the  current density
$j_{\mu}(\omega,q_{x},y)$ was obtained\cite{11} 
when the components of the electric field of the wave, ${\bf E}$,
are smooth on the $\ell_{0}$ scale. This   is not well justified. 
In addition,   
nonlocal effects  and the screening of the 2DEG were neglected.
Later on, within a RPA framework  
we  showed  \cite{12} that the effect of nonlocality and screening
by the 2DEG on the fundamental EMP is small and justifies well,  for weak
dissipation, the results of Ref. \cite{11} for the fundamental EMP.
Here,  using the same  RPA framework we consider
the fundamental EMPs at the edge of a DQW  
for $\nu=2$, when two occupied Landau sublevels (LSLs),
with $n=0$ and the same spin orientation, correspond to
different DQW quantum numbers $\alpha_{z}=-$ and  $\alpha_{z}=+$.
As all studies  we are aware of treat EMPs in  single quantum well systems, 
the present microscopic generalization of the model of
Refs. \cite{10a} and \cite{11} to   EMPs in   DQWs appears amply justified.

In Sec. II we present the one-electron characteristics of a DQW channel,
the wave charge density, inhomogeneous along $y$
on a scale $\sim \ell_{0} \ll \lambda$, and the corresponding electric
potential. Further, we derive the integral equations for
the fundamental EMPs in a DQW
for negligible and  strong tunnel coupling. In Sec. III we %for $\nu=2$
obtain the corresponding dispersion relations of the
fundamental EMPs for very low temperatures.
Finally, in Sec. IV we discuss our theory and make concluding remarks.

\section{BASIC RELATIONS}

\subsection{DQW channel characteristics}

We consider a DQW, of width $W$ and of length $L_x=L$,
in the presence of a strong magnetic field $B$ along the $z$ axis.
In the absence of a self-consistent wave potential
$V({\bf r},t)=V(\omega_{0},q_{x},y,z)
\exp[-i(\omega_{0} t-q_{x} x)]+ c.c.$, we assume that the motion
of an electron in the 2D-plane ($x,y$) is decoupled from it's
motion along the $z$ axis. The latter is described by the
Schrodinger equation
%, i.e., man-made potential growth direction; ${\bf r}=\{x,y,z\}$.
%Along the $z$-direction we have t

        \begin{equation}
        \hat{h}_{z} X_{\alpha_{z}}(z)
        =[-\frac{\hbar^2}{2m^*}\frac{d^2}{dz^2}+V_z]X_{\alpha_{z}}(z)=
        E_{\alpha_{z}} X_{\alpha_{z}}(z).
        \label{1}
        \end{equation}
The DQW consists of  two  square wells along the $z$ axis,
with finite depth $U$, widths $d_{l}$ and $d_{r}$, and separated by a barrier
of width $d$.
In addition, we assume that $m^*$ is independent of $z$. As is usual
in  DQW studies,
we assume that only a pair of tunnel-connected closely located levels,
formed from the ground states of the left  $(l)$ and right $(r)$
{\em isolated} QWs, is essential.
For small barrier penetration \cite{13}  we can use these isolated
$r$ and $l$ well eigenfunctions, denoted,
respectively, by $X_{r}(z)$ and $X_{l}(z)$
as a basis for solving
Eq. ($\ref{1}$). With $ \alpha_{z}=\pm$ the corresponding normalized orbitals
$X_{\alpha_{z}}$ read \cite{13}

        \begin{equation}
        X_+=N[X_l+(2T/(\Delta+\Delta_T))X_r], \;\;\;\;\;
        X_-=N[X_r-(2T/(\Delta+\Delta_T))X_l] .
        \label{2}
        \end{equation}
Here $N=\sqrt{(\Delta+\Delta_T)/2\Delta_T}$ is the normalization
factor, $T=-T_0 \exp(-\kappa d)$
($T_0\approx 2\pi^2\hbar^2/m^*\kappa d_l^{3/2}d_r^{3/2}$) is the
tunneling matrix element, and $\kappa^{-1} \approx \hbar/\sqrt{2m^*U}$.
Further, $\Delta$ is the energy splitting of the
isolated $l$ and $r$ levels, and $\Delta_T=\sqrt{\Delta^2+4T^2}$
is the splitting between the $\alpha_{z}=+$  and  $\alpha_{z}=-$
levels; $\Delta_T$  takes into account tunnel
branch-off. The eigenvalues $E_+$ and $E_-$ are $+\Delta_T/2$ and
$-\Delta_T/2$, respectively. Here weak barrier penetration
means  $\exp(-2\kappa d) \ll 1$ and $\kappa d_{l,r} \gg \pi$.
In the calculations of pertinent matrix elements we can approximate
$X_{l,r}(z)$ by $\sqrt{2/d_{l,r}} \cos(\pi(z-z_{l,r})/d_{l,r})$ for
$|z-z_{l,r}| \leq d_{l,r}/2$, and by 0 for $|z-z_{l,r}| \geq d_{l,r}/2$,
where $z_l$, and $z_r$ are the centers of $l$ and $r$ QWs, respectively.
That is, the matrix elements can be calculated without taking into account
the barrier penetration of the  isolated well orbitals, cf.  Ref. \cite{13}.

The lateral confining potential is taken flat in the interior of the 2DEG
($V_{y}=0$) and parabolic at its edges, e.g., $V_{y} =m^*\Omega^2
(y-y_r)^2/2$, $y\geq y_r$.  $V_{y}$ is assumed smooth on the scale of
$\ell_{0}=(\hbar/m^{*} \omega_{c})^{1/2}$
such that $\Omega \ll \omega_{c}$, where
$\omega_{c}=|e|B/m^* $ is the cyclotron frequency.
We  will also consider the case of a tunnel-decoupled DQW,
with different confining potentials  in the $l$ and $r$
QWs, i.e., $V_{y}^{(l,r)} =m^*\Omega^2 (y-y_{r}^{(l,r)})^2/2$,
$y\geq y_{r}^{(l,r)}$ with $y_{r}^{(l)} \neq y_{r}^{(r)}$.
In the Landau gauge for the vector potential ${\bf A}=(-By, 0, 0)$
and without the interaction $V({\bf r},t)$, the total one-electron
Hamiltonian $\hat{h}^{0}$ is given as
$\hat{h}^{0}=\hat{h}_{2d}+\hat{h}_{z}$, and

        \begin{equation}
        \hat{h}_{2d}=[(\hat{p}_x +eBy)^2 +\hat{p}_{y}^{2}]/2m^* +V_{y}+
        g^{*} \mu_B \hat{S}_z B/2   \;\; .
        \label{3}
        \end{equation}
Here $\hat{{\bf p}}$ is the momentum operator, $g^{*}(<0)$ the
effective Land\'e g-factor, and $\mu_B$ the Bohr
magneton.  $\hat{S}_z$ is the z-component of the spin operator with
eigenvalues $\sigma=1$ and $\sigma=-1$ for spin up ($\uparrow$) and
down ($\downarrow$), respectively. The eigenvalues and eigenfunctions
corresponding to Eq. (\ref{3}) for the right edge of the channel, where
$y_{0} \equiv y_{0}(k_{x})=\ell_{0}^{2} k_{x} \geq y_{r}$,
are well approximated \cite{14}  by

        \begin{equation}
        E_{\alpha_{2d}}\equiv E_{n,\sigma}(k_{x})=
        (n+1/2)\hbar \omega_{c}+m^*\Omega^2 (y_{0}-y_{r})^2/2
        +\sigma g^{*} \mu_B B/2 ,\\
        \label{4}
        \end{equation}
and
        \begin{equation}
        |\alpha_{2d}> \equiv \psi_{\alpha_{2d}}(\vec{\rho}) |\sigma>
        =e^{ik_x x}\Psi_n(y-y_0)|\sigma> /\sqrt{L},\\
        \label{5}
        \end{equation}
respectively.  Here $\vec{\rho}=\{x,y\}$,
$k_{r}=y_{r}/\ell_{0}^{2}$,
 $\alpha_{2d} \equiv \{n,k_x,\sigma\}$,
$\Psi_n(y)$ is a harmonic oscillator function,
% where the renormalized magnetic length
% $\tilde{\ell}=(\hbar/m^* \tilde{\omega})^{1/2} \approx \ell_{0}$,
% $\tilde{\omega}=(\omega_{c}^{2}+\Omega^{2})^{1/2}$,
% cf. with Ref. \cite{14};
$|\sigma>=\psi_{\sigma}(\sigma_1)=\delta_{\sigma \sigma_1}$ is the
spin-wave function, and $\sigma_1=\pm1$.
Then the eigenvalues of $\hat{h}^{0}$ are %given as
$E_{\alpha}=E_{\alpha_{2d}}+E_{\alpha_{z}}$ and its eigenfunctions
$|\alpha>$= %are given as
$\psi_{\alpha_{2d}}(\vec{\rho}) X_{\alpha_{z}}(z) |\sigma>$.

We will consider only the case of $\nu=2$, when only
two tunnel-split levels
%corresponding to $\Delta_T$-splitted n=o Landau level,
are occupied. These Landau sublevels (LSLs)
have the same Landau  and spin quantum numbers,   $n=0$ and $\sigma=1$,
but   different  quantum numbers
$\alpha_{z}=\pm$. In line with Refs. \cite{15} and  \cite{16},
we assume that $\hbar \omega_{c} > \Delta_T$ and $|g^{*}| \mu_B B > \Delta_T$.
Then the one-electron energy spectrum (for $y_{0} \geq y_{r}$) for
the two occupied LSLs is given by

        \begin{equation}
        E_{0,k_x,1,\pm} \equiv E_{0,\pm}(k_x)=
        \Big[\hbar \omega_{c}+m^*\Omega^2 (y_{0}-y_{r})^2
        -|g^{*}| \mu_B B \pm \Delta_T\Big]/2 .\\
        \label{6}
        \end{equation}
Equation (\ref{6}) leads to the group velocity of the
edge states $v_{g0}^{\pm}=\hbar^{-1}
\partial E_{0,\pm}(k_{r}+k_{e}^{0,\pm})/\partial k_{x}=\hbar
\Omega^{2}k_{e}^{0,\pm}/m^{*}\omega_{c}^{2}$ with characteristic wave
vector $k_{e}^{0,\pm}=(\omega_{c}/ \hbar\Omega)
\sqrt{2m^{*}\Delta_{F0}^{\pm}}$, where
$\Delta_{F0}^{\pm}=E_{F}-[\hbar \omega_{c}-|g^{*}| \mu_B B
\pm \Delta_T]/2>0$, and $E_F$ is the Fermi energy. The edge of the LSL is
denoted by
$y_{r0}^{\pm}=y_{r}+\ell_{0}^{2} k_{e}^{0,\pm}=
\ell_{0}^{2} k_{r0}^{\pm}$,
where $k_{r0}^{\pm}=k_{r}+k_{e}^{0,\pm}$,  and $W=2y_{r0}^{-}$.
We can also write $v_{g0}^{\pm}= E_{e0}^{\pm}/B$,
where $E_{e0}^{\pm}=\Omega \sqrt{2m^{*}\Delta_{F0}^{\pm}}/|e|$
is the electric field associated with
the slope of the confining potential $V_{y}$ at $y_{r0}^{\pm}$.

For definiteness, we take the background dielectric constant
$\epsilon$ to be spatially homogeneous.
Assuming $|q_{x}| W \gg 1$, we can consider an EMP along the right
edge of the channel, of the form
$A(\omega, q_{x}, y, z) \exp[-i(\omega t-q_{x} x)]$, totally
independent of the left edge.

\subsection{ Wave charge density and electric potential at the DQW edge}

As in Refs.  \cite{11}, \cite{14},
we assume that without interaction
the one-electron density matrix $\hat{\rho}^{(0)}$ is diagonal, i.e.,
$<\alpha|\hat{\rho}^{(0)}|\beta>=f_{\alpha}\delta_{\alpha\beta}$,
where $f_{\alpha}=1/[1+exp((E_{\alpha}-E_F)/k_BT)]$ is the Fermi-Dirac
function. In  applying  the RPA we follow the self-consistent field approach
which is discussed, e.g., in Ref. \cite{17}.
The one-electron
Hamiltonian is $\hat{H}(t)=\hat{h}^0+V(x, y, z,t)$ in the presence
of the interaction taken as a self-consistent wave potential $V({\bf r},t)$.
Then the equation of motion for the density matrix $\hat{\rho}$ reads

        \begin{equation}
        i\hbar\frac{\partial\hat{\rho}}{\partial t}=[\hat{H}(t),\hat{\rho}]
        -\frac{i\hbar}{\tau}(\hat{\rho}-\hat{\rho}^{(0)}),
        \label{8}
        \end{equation}
where $[,]$ denotes the commutator.  On the right hand side (RHS) of
Eq.  (\ref{8}) we have introduced phenomenologically the
infinitesimal term $\propto 1/\tau $ ($\tau \rightarrow \infty$) that
leads to correct
rules for passing an integration contour around the pole
singularities, etc., cf.  Refs.  \cite{17}-\cite{18}.  Notice that
$\tau \rightarrow \infty$ corresponds to the collisionless case
whereas a finite $\tau$ provides the possibility of estimating very
roughly the influence of collisions.  The most effective scattering is
related to intra-LSL and intra-edge transitions
\cite{11}, \cite{14}.

   Applying the Laplace transformation over time to Eq. (\ref{8}) and
writing $\hat{R}(\omega)=\int_{0}^{\infty}e^{i\omega t} \hat{\rho} dt$
and $R_{\alpha \beta}(\omega)=<\alpha|\hat{R}(\omega)|\beta>$
we can present the solution of Eq. (\ref{8}) as a power
series in $V$

        \begin{equation}
        R_{\alpha \beta}(\omega)=\sum_{m=0}^{\infty}
        R_{\alpha \beta}^{(m)}(\omega) ,
        \label{9}
        \end{equation}
where $R_{\alpha \beta}^{(0)}(\omega)=(i f_{\alpha}/\omega)
\delta_{\alpha \beta}$. Because we consider linear EMP's, it is
sufficient  to take into account only the first two
terms $m=0$ and $m=1$ on the RHS of Eq. (\ref{9}).
Then in $V(x,y,z,t)$ we can consider  only  the term
$V(\omega_{0},q_{x},y,z) \exp[-i(\omega_{0} t-q_{x} x)]$, which leads to

        \begin{equation}
        R_{\alpha \beta}^{(1)}(\omega)=\frac{i(f_{\beta}-f_{\alpha})
        <\alpha|V(\omega_{0},q_{x},y,z) e^{iq_{x} x}|\beta>}
        {(\omega-\omega_{0}) [E_{\beta}-E_{\alpha}+\hbar \omega+i\hbar/\tau]} .
        \label{10}
        \end{equation}
Because $V({\bf r},t)$ conserves spin the spin index in
Eq. (\ref{10}) is the same, $\sigma_{\alpha}=\sigma_{\beta}=1$, and
therefore it can be dropped. For simplicity we will
use the notation $\alpha=\{n,k_{x\alpha},\alpha_{z}\}$. Then
taking the trace of the density matrix $\hat{\rho}$ with the electron
charge density operator, $e \delta({\bf r}-\hat{{\bf r}})$, we obtain
the wave charge density as

        \begin{equation}
        \delta \rho(t,x,y,z) \equiv \rho(t,x,y,z)=\frac{e}{2\pi}
        \int_{-\infty+i\eta}^{\infty+i\eta} d \omega e^{-i\omega t}
        \sum_{\alpha \beta} R_{\alpha \beta}^{(1)}(\omega)
        \psi_{\beta}^{*}({\bf r}) \psi_{\alpha}({\bf r}) ,
        \label{11}
        \end{equation}
where $\eta>0$ and $\psi_{\alpha}({\bf r})=
\psi_{\alpha_{2d}}(\vec{\rho}) X_{\alpha_{z}}(z)$.  From Eqs.
(\ref{10}) and (\ref{11}) it follows that
$\rho(t,x,y,z)=\rho(t,q_{x},y,z) \exp(iq_{x} x)$.  Moreover, for
$t/\tau \gg 1$, when contributions related to transitional processes
are negligible,   Eqs.  (\ref{10}) and (\ref{11}) lead to
$\rho(t,q_{x},y,z)=\rho(\omega_{0},q_{x},y,z) \exp(-i\omega_{0}t)$.

To simplify the calculations  we assume that $d_{l} \rightarrow 0$ and $d_{r}
\rightarrow 0$,
i.e., we  neglect the thickness of both QWs, cf.  Ref. \cite{15}.
This leads to $X_{l,r}^{2}(z) \rightarrow \delta (z \pm d/2)$
for $d_{l,r} \rightarrow 0$. Then  Poisson's equation  gives the
wave electric potential  $\phi(t, q_{x},y,z)$ induced by the density
$\rho(t,q_{x},y,z)=\rho^{(l)}(t,q_{x},y) \delta (z+d/2)+
\rho^{(r)}(t,q_{x},y) \delta (z-d/2)$  in the $l$ QW ($z=-d/2$) and
$r$ QW ($z=d/2$) as

         \begin{equation}
        \phi(t, q_{x},y,z=-d/2)=\phi_{l}(t, q_{x},y)=
        \frac{2}{\epsilon}\int_{-\infty}^{\infty}
        dy' K_{lr}(y, y'; t, q_x),
        \label{12}
        \end{equation}
and

        \begin{equation}
        \phi(t, q_{x},y,z=d/2)=\phi_{r}(t, q_{x},y)=
        \frac{2}{\epsilon}\int_{-\infty}^{\infty}
        dy'  K_{rl}(y, y'; t, q_x),
        \label{13}
        \end{equation}
where

        \begin{equation}
        K_{lr}(y, y'; t, q_{x}) =K_{0}(|q_{x}||y-y'|)\rho^{(l)}(t, q_{x},y')+
        K_{0}(|q_{x}| d_{yy'})\rho^{(r)}( t, q_{x}, y').
        \label{13a}
         \end{equation}
Here $ K_{0}(x)$ is the modified Bessel function and
$d_{yy'}=\sqrt{(y-y')^{2}+d^{2}}$.

\subsection{ Integral equations for EMP's }

For $t/\tau \gg 1$, we can change $\rho^{(j)}(t, q_{x},y')$ to
$\rho^{(j)}(\omega_{0},q_{x},y')$ ($j=l,r$) on the RHS of
Eqs.  (\ref{12}) and (\ref{13}).  Then, omitting the common factor
$\exp(-i\omega_{0}t)$, we should
change $\phi_{l,r}(t, q_{x},y)$ to $\phi_{l,r}(\omega_{0}, q_{x},y)$.  In
the absence of an external potential we have $V(\omega_{0},q_{x},y,z)=e
\phi(\omega_{0}, q_{x},y,z)$.
Then using Eqs.  (\ref{11})-(\ref{13}) and integrating Eq.  (\ref{11})
over $z$ once in the neighborhood  of $z=-d/2$ and once in
that of $z=d/2$, we obtain a
system of two integral  equations for $\rho^{(l)}(\omega_{0}, q_{x},y)$ and
$\rho^{(r)}(\omega_{0}, q_{x},y)$; if we  omit the subscript $0$ in
$\omega_{0}$ they read

        \begin{eqnarray}
        \nonumber
        \rho^{(j)}(\omega, q_{x}&,&y)=\frac{2e^{2}}{\epsilon L}
        \sum_{n_{\alpha},n_{\beta}=0}^{\infty} \sum_{k_{x\alpha}}
        \sum_{\alpha_{z},\beta_{z}}
\frac{(f_{\beta}-f_{\alpha})<\alpha_{z}|\beta_{z}>_{j}}
{E_{\beta}-E_{\alpha}+\hbar \omega+i \hbar/\tau}\
\Pi_{n_{\alpha}n_{\beta}}( y, k_{x\alpha}, k_{x\beta} ) \\*
        \nonumber
        \ \\
&&\times\int_{-\infty}^{\infty} d\tilde{y}  \int_{-\infty}^{\infty} dy'
\ \Pi_{n_{\alpha}n_{\beta}}(\tilde{y}, k_{x\alpha}, k_{x\beta})
\{[<\alpha_{z}|\beta_{z}>_{l}  K_{lr}(\tilde{y}, y'; \omega, q_x)]+
[r \leftrightarrow l]\};
        \label{14}
        \end{eqnarray}
where
$\Pi_{n_{\alpha}n_{\beta}}( y, k_{x\alpha}, k_{x\beta})=
\Psi_{n_{\alpha}}(y-y_0(k_{x\alpha})) \Psi_{n_{\beta}}(y-y_0(k_{x\beta}))$,
$k_{x\beta}=k_{x\alpha}-q_{x}$,
$<\alpha_{z}|\beta_{z}>_{l}=\int_{-d/2-0}^{-d/2+0} dz X_{\alpha_{z}}
X_{\beta_{z}}$ and $<\alpha_{z}|\beta_{z}>_{r}=
\int_{d/2-0}^{d/2+0} dz X_{\alpha_{z}} X_{\beta_{z}}$.

  The integral equations (\ref{14}) take into account the nonlocality in
  the electron current density $\propto \int dy' \sigma_{\mu
  \nu}(y,y') E_{\nu}(y')$, and the screening by the edge and bulk
  states of the 2DEG.  In Ref.  \cite{19} it is shown that screening by the
  edge states can be strong.  We consider very low temperatures $T$
  satisfying $\hbar v_{g0}^{\pm} \gg \ell_{0} k_{B} T$ and use the
  long-wave length limit $q_{x} \ell_{0} \ll 1$.  Further we will obtain
  only the low-frequency solutions of Eqs.  (\ref{14})  for i)
  negligibly small tunneling $\omega \ll \omega_{c}$ and ii) strong
  tunneling $\omega \ll \Delta_T/\hbar$.  In both cases we assume
  $\omega_{c}>\Delta_T/\hbar$.

%Then for low frequency regime in (ii) c
{\it Case ii)}. Comparing the terms $\propto
f_{\beta^{*}}$ on  the RHS of Eq.  (\ref{14}) , for given
$n_{\beta^{*}}=0$ and $\beta_{z}^{*}=+$ or
$-$, we find that the contribution to the
sum over $n_{\alpha}$ and $\alpha_{z}$ with $n_{\alpha}=n_{\beta^{*}}=0$ and
$\alpha_{z}=\beta_{z}^{*}$, is much  larger than that of any other term in
this sum (e.g.,  the term with $n_{\alpha}=0$ and $\alpha_{z} \neq
\beta_{z}^{*}$)  or that  of all other terms with $n_{\alpha}
> 0$.  The small parameter is  $\hbar |\omega-q_{x}v_{g
0}^{\pm}(k_{x \beta})|/\Delta_T \ll 1$, where $v_{g 0}^{\pm}(k_{x
\beta})=\hbar^{-1} \partial E_{0,\pm}(k_{x \beta})/\partial k_{x
\beta}$ is the group velocity of any occupied state $\{ 0,k_{x
\beta},\beta_{z}^{*} \}$ of the $\beta_{z}^{*}$ LSL.  We assume that both
occupied LSLs have one intersection with the Fermi level at the edge
of the channel and denote the group velocity of its edge states by
$v_{g0}^{\pm} \equiv v_{g0}^{\pm}(k_{r0}^{\pm})$.  The
small parameter  given above  implies
$\hbar q_{x} v_{g0}^{-}/\Delta_T \ll 1$,
since $v_{g0}^{-}$ is typically  larger than  $v_{g0}^{+}$.
Similar results hold for the terms $\propto
f_{\alpha^{*}}$ in the sum over $n_{\beta}$ and $ \beta_{z}$.
Hence for $\hbar \omega \ll \Delta_T$ and $\hbar q_{x} v_{g
0}^{\pm} \ll \Delta_T$
the terms with $n_{\alpha} \neq
n_{\beta}$ or/and $\alpha_{z} \neq \beta_{z}$ can be neglected. This
leads to a system  of simpler integral equations for $\rho^{(l)}(\omega,
q_{x},y)$ and $\rho^{(r)}(\omega, q_{x},y)$

        \begin{eqnarray}
        \nonumber
        \rho^{(j)}(\omega &,& q_{x},y)=\frac{2e^{2}}{\epsilon L}
        \sum_{\alpha_{z}, k_{x\alpha}}  %\sum_{\alpha_{z}}
        \frac{<\alpha_{z}|\alpha_{z}>_{j} (f_{0,k_{x\alpha}-q_{x},\alpha_{z}}-
        f_{0,k_{x\alpha},\alpha_{z}})}
        {E_{0,\alpha_{z}}(k_{x\alpha}-q_{x})-E_{0,\alpha_{z}}(k_{x\alpha})
        +\hbar \omega+i \hbar/\tau}\
        \Pi_{0 0}(y, k_{x\alpha}, k_{x\alpha}-q_x)
        \nonumber
        \ \\
        \nonumber
        \ \\
        &&\times %\Psi_{0}(y-y_0(k_{x\alpha}-q_{x}))
        \int_{-\infty}^{\infty} d\tilde{y}  \int_{-\infty}^{\infty} dy'
        \ \Pi_{00}(\tilde{y}, k_{x\alpha}, k_{x\alpha}-q_x)
        \Big \{[<\alpha_{z}|\alpha_{z}>_{l}K_{lr}(\tilde{y}, y'; \omega, q_x)]
        + [r \leftrightarrow l]\Big \}.
        \label{15}
        \end{eqnarray}

{\it Case i)}. For  negligibly small tunneling we have
$2|T|/\Delta \ll 1$ and from Eq. (\ref{2}) it follows that
$X_{+}(z) \approx X_{l}(z)$ and $X_{-}(z) \approx X_{r}(z)$. Again we
can take $n_{\alpha}=n_{\beta}=0$. Then  Eq. (\ref{14})
 for $\omega \ll \omega_{c}$, $q_{x} v_{g 0}^{\pm} \ll \omega_{c}$,
$j=l$, and  $\alpha_{z}=\beta_{z}=+$ gives

        \begin{eqnarray}
        \nonumber
        \rho^{(l)}(\omega, q_{x}&,&y)=\frac{e^{2}}{\pi \epsilon}
        \int_{-\infty}^{\infty} dk_{x\alpha}
        \ \frac{f_{0,k_{x\alpha}-q_{x},+}-f_{0,k_{x\alpha},+}}
        {E_{0,+}(k_{x\alpha}-q_{x})-E_{0,+}(k_{x\alpha})
+\hbar \omega+i \hbar/\tau}\ \Pi_{00}( y, k_{x\alpha}, k_{x\alpha}-q_x)\\*
        \nonumber
        \ \\
        %\nonumber
        &&\times
        \int_{-\infty}^{\infty} d\tilde{y}  \int_{-\infty}^{\infty} dy'
        \ \Pi_{00}(\tilde{y}, k_{x\alpha}, k_{x\alpha}-q_x)
        \ K_{lr}(\tilde{y}, y'; \omega, q_x) \; ;
        \label{16}
        \end{eqnarray}
for $j=r$ and $\alpha_{z}=\beta_{z}=-$ the result is

        \begin{eqnarray}
        \nonumber
         \rho^{(r)}(\omega, q_{x},y)&=&\frac{e^{2}}{\pi \epsilon}
        \int_{-\infty}^{\infty} dk_{x\alpha}
        \ \frac{f_{0,k_{x\alpha}-q_{x},-}-f_{0,k_{x\alpha},-}}
        {E_{0,-}(k_{x\alpha}-q_{x})-E_{0,-}(k_{x\alpha})
        +\hbar \omega+i \hbar/\tau}\ \Pi_{00}( y, k_{x\alpha},
k_{x\alpha}-q_x)\\*
        \nonumber
        \ \\
        %\nonumber
   &&\times \int_{-\infty}^{\infty} d\tilde{y}  \int_{-\infty}^{\infty} dy'
        \ \Pi_{00}(\tilde{y}, k_{x\alpha}, k_{x\alpha}-q_x)
        \ K_{rl}(\tilde{y}, y'; \omega, q_x).
        \label{17}
        \end{eqnarray}
We will study only the fundamental EMPs  following from Eqs.
(\ref{15})-(\ref{17}).

\section{Fundamental EMPs, $\nu=2$}

\subsection{ Negligibly small tunnel coupling}

We now consider case (i), when $2|T|/\Delta \ll 1$ holds, and we have the
system of
Eqs. (\ref{16}) and (\ref{17}) for $\rho^{(l)}(\omega, q_{x},y)$ and
$\rho^{(r)}(\omega, q_{x},y)$ coupled only by the Coulomb interaction.
In the assumed long-wavelength limit $q_{x} \ell_{0} \ll 1$ we can
approximate $f_{0,k_{x\alpha}-q_{x},\pm}-f_{0,k_{x\alpha},\pm}$ by
$q_{x} \delta(k_{x\alpha}-k_{r0}^{\pm})$ %in Eqs.  (\ref{16}), (\ref{17})
and neglect the small shift $\propto q_{x} \ell_{0}$ in the
argument of   $\Psi_{0}(x-y_0(k_{x\alpha})+\ell_{0}^{2} q_{x})$, i.e.,
we can approximate the latter by $\Psi_{0}(x-y_0(k_{x\alpha}))$.  Then
from Eqs.  (\ref{16}) and (\ref{17}), after integration over
$k_{x\alpha}$, we obtain

        \begin{equation}
        %\nonumber
        \rho^{(l)}(\omega, q_{x},y)=\frac{e^{2}}{\pi \hbar \epsilon}
   \ \frac{q_{x}\Psi_{0}^{2}(y-y_{r0}^{+})}{\omega-q_{x} v_{g0}^{+}+i/\tau}
        %\Psi_{0}^{2}(y-y_{r0}^{+})
        \int_{-\infty}^{\infty} d\tilde{y}  \int_{-\infty}^{\infty} dy'
        \ \Psi_{0}^{2}(\tilde{y}-y_{r0}^{+}) \ K_{lr}(\tilde{y}, y'; \omega,
q_x)
        \label{18}
        \end{equation}
and

        \begin{equation}
        %\nonumber
        \rho^{(r)}(\omega, q_{x},y)=\frac{e^{2}}{\pi \hbar \epsilon}
        \ \frac{q_{x}\Psi_{0}^{2}(y-y_{r0}^{-})}{\omega-q_{x} v_{g0}^{-}+i/\tau}
        %\Psi_{0}^{2}(y-y_{r0}^{-})
        \int_{-\infty}^{\infty} d\tilde{y}  \int_{-\infty}^{\infty} dy'
        \ \Psi_{0}^{2}(\tilde{y}-y_{r0}^{-}) \ K_{rl}(\tilde{y}, y'; \omega,
q_x) ,
        \label{19}
        \end{equation}
respectively. The  solution   of Eqs. (\ref{18}) and  (\ref{19}) can be
written as

        \begin{equation}
        \rho^{(l)}(\omega, q_{x},y)=\rho^{(l)}(\omega, q_{x})
        \ \Psi_{0}^{2}(y-y_{r0}^{+}) , \;\;\;
        \rho^{(r)}(\omega, q_{x},y)=\rho^{(r)}(\omega, q_{x})
        \ \Psi_{0}^{2}(y-y_{r0}^{-}) .
        \label{20}
        \end{equation}
Since $X_{+}(z) \approx X_{l}(z)$ and
$X_{-}(z) \approx X_{r}(z)$, we can consider $v_{g0}^{+}$ and $y_{r0}^{+}$
as parameters of the $l$ QW  that are independent of those of the
$r$ QW  $v_{g0}^{-}$ ($y_{r0}^{-}$) ; the same
holds for $E_{0,+}(k_{x})$ and $E_{0,-}(k_{x})$. Thus, in
Eqs. (\ref{18})-(\ref{20}) we can take, e.g.,
$\Delta y=y_{r0}^{-}-y_{r0}^{+} \gg \ell_{0}$.
In particular, here we can  assume different confining
potentials  in the $l$ and $r$ QWs, $V_{y}^{(l,r)}$.

Substituting Eq. (\ref{20}) into Eqs. (\ref{18}),(\ref{19}) gives
a system of two linear homogeneous  equations  for the unknowns
$\rho^{(l,r)}(\omega, q_{x})$.  The condition of a nontrivial
solution  gives the dispersion relation (DR) for the renormalized, by
the Coulomb coupling,  single-well fundamental EMP
($\omega_{\alpha_{z}}^{(c)}(q_{x})\equiv \omega_{\alpha_{z}}^{(c)}$) as

        \begin{equation}
        \omega_{\alpha_{z}}^{(c)} =
         q_{x}V_{g+} -
        \frac{i}{\tau}+q_{x} a_{00}(q_{x})\ \frac{e^{2}}{\pi \hbar \epsilon}
         -\alpha_{z} q_{x}
         [(\frac{e^{2}}{\pi \hbar \epsilon})^{2}
        \ b_{00}^{2}(q_{x},\Delta y,d)+
        V_{g-}^{2}]^{1/2} ,
        \label{21}
        \end{equation}
where $V_{g\pm}=(v_{g0}^{-}\pm v_{g0}^{+})/2$ and
$\alpha_{z}=\pm$ corresponds to the  $\pm$ LSL. Here

        \begin{equation}
        b_{00}(q_{x},\Delta y,d)=\int_{-\infty}^{\infty} dx\
        \int_{-\infty}^{\infty} dx'\ \Psi_{0}^{2}(x) \ \Psi_{0}^{2}(x')
        \ K_{0}(|q_{x}|\sqrt{[(x-x')+\Delta y]^{2}+d^{2}})
        \label{22}
        \end{equation}
and $a_{00}(q_{x})=b_{00}(q_{x},0,0) \approx
[\ln(1/|q_{x}|\ell_{0})+3/4]$ \cite{11}; it is assumed that
$v_{g0}^{-} \geq v_{g0}^{+} >0$.  Without inter-QW Coulomb coupling,
i.e., with $b_{00}(q_{x},\Delta y,d) = 0$ in Eq.
(\ref{21}), we obtain the DR of the fundamental EMPs
for totally {\it  decoupled } QWs as
$\omega_{-}^{(r)} =q_{x}v_{g0}^{-}+ [e^{2}/\pi \hbar \epsilon]
q_{x} a_{00}(q_{x})-i/\tau$, for the $r$ QW, and  as
$\omega_{+}^{(l)} =q_{x}v_{g0}^{+}+ [e^{2}/\pi \hbar \epsilon]
q_{x} a_{00}(q_{x})-i/\tau$, for the $l$ QW.  If we neglect  dissipation,
this result  coincides with that of Ref.  \cite{11} for $\nu=1$ in a
single QW.  For a GaAs-based DQW ($\epsilon=12.5$) we have
$2e^{2}/\pi \hbar \epsilon \approx 10^{7}$ cm/sec which is essentially larger
than the typical $v_{g0} \alt 10^{6}$ cm/sec , cf.  \cite{20},
\cite{11}. Moreover, for $D=(\Delta y^{2}+d^{2})^{1/2} \leq 0.4/q_{x}
\gg \ell_{0}$ it follows that $b_{00}(q_{x},\Delta y,d) > 1$.  Then
for $q_{x} D \ll 1$ and $D^{2}/\ell_{0}^{2} \gg 1$   Eq.  (\ref{21}) gives
 the DR of the renormalized fundamental EMP of
the $r$ QW, i.e., the {\it fast} EMP, as

        \begin{equation}
        \omega_{-}^{(c)}=q_{x} V_{g+} +q_{x}\
        \frac{e^{2}}{\pi \hbar \epsilon}
        \ [2 \ln(\frac{1}{|q_{x}|\ell_{0}})-
        \ln(\frac{D}{\ell_{0}})+0.85]-\frac{i}{\tau} ;
        \label{23}
        \end{equation}
and that of the  $l$ QW, i.e.,  the {\it slow}  EMP, as

        \begin{equation}
        \omega_{+}^{(c)}= q_{x}V_{g+}+q_{x}\
        \frac{e^{2}}{\pi \hbar \epsilon}
        \ [\ln(\frac{D}{\ell_{0}})+0.65]-\frac{i}{\tau}.
        \label{24}
        \end{equation}
Substituting  $\omega_{-}^{(c)} $ and $\omega_{+}^{(c)} $, given by
Eqs.  (\ref{23}) and (\ref{24}), into Eq.  (\ref{18}) we obtain
$\rho^{(l)}(\omega_{-}^{(c)} , q_{x})/
\rho^{(r)}(\omega_{-}^{(c)} , q_{x}) \approx 1$ and
$\rho^{(l)}(\omega_{+}^{(c)} , q_{x})/
\rho^{(r)}(\omega_{+}^{(c)} , q_{x}) \approx -1$, respectively.
For $v_{g0}^{-}=v_{g0}^{+}$ these ratios are exactly  $1$ and $-1$,
respectively. Then, as can be seen from Eq. (\ref{20}), the renormalized
fundamental EMP of the $r$ QW, Eq. (\ref{23}),
shows an {\it acoustic} spatial behavior in the $y$ direction and
that of the $l$ QW,  Eq. (\ref{24}),  an
{\it optical} one  though its DR  is purely {\it acoustic}.

For negligible dissipation the DRs (\ref{23}) and (\ref{24})
are shown in Fig.  1  by the solid ($D^{2}/\ell_{0}^{2}=3$)
and dashed ($D^{2}/\ell_{0}^{2}=10$) curves, respectively,
for $v_{g0}^{-} \approx v_{g0}^{+}=10^{6}$
cm/sec, $\epsilon=12.5$, and  $\omega_{*}=e^{2}/\pi \hbar \epsilon
\ell_{0}$. The dotted curve shows the DR
of an isolated, single-well fundamental EMP
$\omega_{-}^{(l)} = \omega_{+}^{(r)} =q_{x}v_{g0}^{-}+[e^{2}/
\pi \hbar \epsilon]
q_{x} [\ln(1/(|q_{x}|\ell_{0}))+3/4]$.  The upper solid and dashed
curves represent the fast  EMP and the lower ones the  slow   EMP.

\subsection{ Strong tunnel coupling}

We now  consider   case  (ii) in which
$2|T| \agt \Delta$ holds.  As in Sec. III A,  we make the approximations
$f_{0,k_{x\alpha}-q_{x},\pm}-f_{0,k_{x\alpha},\pm}\approx
q_{x} \delta(k_{x\alpha}-k_{r0}^{\pm})$ and
$\Psi_{0}(x-y_0(k_{x\alpha})+\ell_{0}^{2} q_{x}) \approx
\Psi_{0}(x-y_0(k_{x\alpha}))$. Then  integration over
$k_{x\alpha}$ in  Eq. (\ref{15}) results in

        \begin{eqnarray}
        \nonumber
        \rho^{(j)}(\omega, q_{x}&,&y)=q_{x}\ \frac{e^{2}}{\pi \hbar \epsilon}
        \sum_{\alpha_{z}=\pm}
        \frac{<\alpha_{z}|\alpha_{z}>_{j}}
        {\omega-q_{x}v_{g0}^{\alpha_{z}}+i/\tau}
        \ \Psi_{0}^{2}(y-y_{r0}^{\alpha_{z}})\\*
        \nonumber
        \ \\
        %\nonumber
        && \times\int_{-\infty}^{\infty} d\tilde{y}
        \int_{-\infty}^{\infty} dy'
        \ \Psi_{0}^{2}(\tilde{y}-y_{r0}^{\alpha_{z}})
         \{[<\alpha_{z}|\alpha_{z}>_{l}
        K_{lr}(\tilde{y}, y'; \omega, q_x)] +
        [r \leftrightarrow l]\} ,
        \label{25}
        \end{eqnarray}
where $j=l, r$. The solution of Eqs. (\ref{25}) can be written as

        \begin{equation}
        \rho^{(i)}(\omega, q_{x},y)=\sum_{\alpha_{z}=\pm}
        <\alpha_{z}|\alpha_{z}>_{i} \rho_{\alpha_{z}}(\omega, q_{x})
        \ \Psi_{0}^{2}(y-y_{r0}^{\alpha_{z}}) ,
        \label{26}
        \end{equation}
where $i=l,r$. Substituting Eq. (\ref{26}) into Eq.  (\ref{25})
and demanding that the  factors in front of  $\Psi_{0}^{2}(y-y_{r0}^{\pm})$
on both sides of Eq. (\ref{25}) be equal  for $j=l$ and
for $j=r$, leads to  a system of two  linear  equations
for $\rho_{\pm}(\omega, q_{x})$ written as
        \begin{eqnarray}
        \nonumber
        &&\rho_{\alpha_{z}}(\omega,q_{x}) \Big\{ \frac{\pi \hbar \epsilon}
        {e^{2} q_{x}} (\omega-q_{x}v_{g0}^{\alpha_{z}}+i/\tau)- N^{4}
        [(1+M^{-8}) a_{00}(q_{x})+ 2M^{-4} b_{00}(q_{x},0,d)] \Big\}\\*
        \nonumber
        \ \\
        &&-\rho_{-\alpha_{z}}(\omega,q_{x})
        N^{4}  \Big\{ [1+M^{-8}] b_{00}(q_{x},\Delta y,d)+
        2M^{-4} b_{00}(q_{x},\Delta y,0) \Big\}=0.
        \label{27}
        \end{eqnarray}
Here $N^2= (\Delta+\Delta_{T})/2\Delta_{T},\
M^2= (\Delta+\Delta_{T})/2T,
\alpha_{z}=-$
for the first  equation and $\alpha_{z}=+$
for the second one. The  condition for  a
nontrivial solution of these equations gives the DR
of the renormalized, due to tunneling  and   Coulomb coupling,
 single-well fundamental EMP
        \begin{eqnarray}
        \nonumber
        \omega_{\alpha_{z}}^{(tc)} &= & q_{x}V_{g+}-
        \frac{i}{\tau}+
        q_{x} \
        \frac{e^{2}}{\pi \hbar \epsilon}
        N^{4}
        \Big\{ [1+M^{-8}] a_{00}(q_{x})+ 2M^{-4} b_{00}(q_{x},0,d)\Big\}\\*
        \nonumber
        \ \\
        &&-\alpha_{z} q_{x}  \Big\{ (\frac{e^{2}}{\pi \hbar \epsilon})^{2}
        N^{8} [(1+M^{-8})   b_{00}(q_{x},\Delta y,d)+
        2M^{-4} b_{00}(q_{x},\Delta y,0) ]^{2}+ V_{g-}^{2}\Big\}^{1/2} .
        \label{28}
        \end{eqnarray}
Notice that for finite $\Delta$ and $T \rightarrow 0$, i.e., without
tunnel coupling,  Eq. (\ref{28}) coincides with  Eq. (\ref{21}).
Further, the first term under the square root in  Eq. (\ref{28}) is
typically much larger %(cf. with discussion of  Eq. (\ref{21}))
than the second one. Then

        \begin{eqnarray}
        \nonumber
        \omega_{\alpha_{z}}^{(tc)} =&&
         q_{x}V_{g+}-\frac{i}{\tau}
        + q_{x}\ \frac{e^{2}}{\pi \hbar \epsilon}
        \Big[a_{00}(q_{x})-\alpha_{z} b_{00}(q_{x},\Delta y,d)\Big]\\*
        \nonumber
        \ \\
        %\nonumber
 &&- q_{x}\ \frac{2e^{2}}{\pi \hbar \epsilon} \ \frac{T^2}{\Delta_{T}^2}
        \Big[a_{00}(q_{x})-b_{00}(q_{x},0,d)\Big];
        \label{29}
        \end{eqnarray}
on the RHS we have neglected the cotribution
$\alpha_{z}[b_{00}(q_{x},\Delta y,0)-b_{00}(q_{x},\Delta y,d)]$, which
is small in comparison with $[a_{00}(q_{x})-b_{00}(q_{x},0,d)]$ due to
the assumed condition $\Delta y^{2}/d^{2} \gg 1$.  Now in the actual
region $1 \geq d/\ell_{0} \geq 1/4$ we numerically evaluate
$[a_{00}(q_{x})-b_{00}(q_{x},0,d)] \approx d/\ell_{0}$ with relative
precision $\leq 10 \%$.  Then Eq.  (\ref{29}) gives the DR
 of the {\it fast}  EMP as

        \begin {equation}
        %\nonumber
        \omega_{-}^{(tc)} =
         q_{x}V_{g+}
        +q_{x}\ \frac{e^{2}}{\pi \hbar \epsilon}
        \ [2 \ln(\frac{1}{|q_{x}|\ell_{0}})-
        \ln(\frac{D}{\ell_{0}})+0.85
         -\frac{2 d}{ \ell_{0}}
        \ \frac{T^{2}}{\Delta^{2}+4T^{2}}]-\frac{i}{\tau} ,
        \label{30}
        \end{equation}
and that  of the {\it slow }  EMP as

        \begin{equation}
        \omega_{+}^{(tc)} =  q_{x}V_{g+}
        +q_{x}\ \frac{e^{2}}{\pi \hbar \epsilon}
        \ [\ln(\frac{D}{\ell_{0}})+0.65
         -\frac{2d}{ \ell_{0}}
        \ \frac{T^{2}}{\Delta^{2}+4T^{2}}]-
          \frac{i}{\tau} ,
        \label{31}
        \end{equation}
where $\Delta y/\ell_{0}= \ell_{0}
[k_{e}^{0,-}-k_{e}^{0,+}] \approx (\omega_{c}/\Omega)
[(\Delta^{2}+4T^{2})/2\hbar \omega_{c} \Delta_{F0}]^{1/2}$ for
 $\Delta_{F0}^{\pm}=\Delta_{F0} \mp \Delta_{T}/2$ and
$\Delta_{F0} \gg \Delta_{T}/2$.
Using  Eqs. (\ref{30}), (\ref{31}), and (\ref{27}) we obtain
 $\rho_{-}(\omega_{-}^{(tc)} ,q_{x})/
\rho_{+}(\omega_{-}^{(tc)},q_{x})=1$ for the fast  mode
and  $\rho_{-}(\omega_{+}^{(tc)} ,q_{x})/
\rho_{+}(\omega_{+}^{(tc)} ,q_{x})=-1$ for the slow  mode.
The corresponding charge density profiles are obtained from Eq. (\ref{26})
 as

\begin{eqnarray}
\nonumber
&&\frac{\rho^{(l)}(\omega_{\mp}^{(tc)}, q_{x},y)}{
\rho_{-}(\omega_{\mp}^{(tc)}, q_{x})}=
N^{2} \{
M^{-4} \Psi_{0}^{2}(y-y_{r0}^{-}) \pm
\Psi_{0}^{2}(y-y_{r0}^{+}) \} \;\; ,\\*
\nonumber
\ \\
%\nonumber
&&\frac{\rho^{(r)}(\omega_{\mp}^{(tc)}, q_{x},y)}{
\rho_{-}(\omega_{\mp}^{(tc)}, q_{x})}=
N^{2} \{ \Psi_{0}^{2}(y-y_{r0}^{-}) \pm
M^{-4} \Psi_{0}^{2}(y-y_{r0}^{+}) \},
\label{32}
\end{eqnarray}
where the upper (lower) sign corresponds to the fast (slow) mode.

For strong tunnel coupling $\Delta \alt 2|T|=1$meV we plot in Fig.  2
the group velocities of both DQW fundamental EMPs, obtained from
Eqs.  (\ref{30}) and (\ref{31}), as function of $\Delta$ with
$v_{*}=e^{2}/\pi \hbar \epsilon$ and $\Delta_{*}=\sqrt{3}$meV.
$\Delta$ can be changed experimentally by a transverse electric field
applied to the DQW structure.  Notice that the group velocity of the fast
mode, $\partial \omega_{-}^{(tc)} / \partial q_{x}$, depends on
$q_{x}$ whereas that of the slow mode, $\partial
\omega_{+}^{(tc)}(q_{x})/ \partial q_{x}$, does not.  For a
GaAs-based DQW, $\Omega=7.8 \times 10^{11} \;\;$sec$^{-1}$ \cite{20},
and $B=9 \;$Tesla we have $\omega_{c}/\Omega \approx 30$, and
$\Delta_{F0}=4$meV.  The value of $d/\ell_{0}$ is $0.5$ for the solid
and dotted curves and $1$ for the dashed curves.  The upper solid and
dashed curves represent the group velocity of the fast
EMP for $q_{x} \ell_{0}=10^{-3}$ and the lower ones that of the slow
EMP.  The dotted curve represents the group velocity
of the fast DQW fundamental EMP for $q_{x} \ell_{0}=10^{-2}$.
For $\Delta=0$  we have $\Delta y/\ell_{0} \approx 2.7$
($\Delta_{T}/2\Delta_{F0}=1/8 \ll 1$) and $\Delta y/\ell_{0} \approx 5.4$
for $\Delta=\sqrt{3}\ $meV
($\Delta_{T}/2\Delta_{F0}=1/4 \ll
1$). Thus the assumed conditions $(\Delta y/d)^{2} \gg 1$, etc.,
are well satisfied in Fig.  2.  In addition, we have $v_{g0}^{\pm}
\approx 5 \times 10^{5}[1 \mp \Delta_{T}/(4\Delta_{F0})] \;\;$cm/sec
and the term $q_{x}(v_{g0}^{-}+v_{g0}^{+})/2$ is independent of
$\Delta$ or $T$.  As is clear from Fig.  2, the change of
$\Delta/\Delta_{*}$ from $0$ to $1$ changes the group velocities of
both DQW fundamental EMPs appreciably %essentially
and especially that of the slow
one which shows a $1.78$ times increase of its group velocity within
that region.  Finally, for larger $d/\ell_{0}$ the group velocity of
the slow mode shows a stronger dependence on $\Delta$ (for small
$\Delta$) whereas that of the fast mode  shows a weaker one.

\section{Discussion and concluding remarks}

The above treatment has shown that in a typical DQW the Coulomb and
tunnel coupling lead to two new DQW fundamental EMPs whose dispersion
laws and spatial structure, along the $y$-axis, are essentially
different than those of single-well fundamental EMPs.
These single-well EMPs
follow from the former if we neglect inter-well coupling.  The fast
 EMP behaves as $\omega \propto q_{x} \ln(1/q_{x})$
whereas the slow  EMP behaves as $\omega \propto
q_{x}$.  The latter has an {\it acoustic} dispersion law though its
spatial structure, along the $y$ axis,  is {\it optical}.  Indeed,
 for both negligible and strong tunneling, when the
 DQW levels are in resonance and
$\Delta$ vanishes, the charge localized at the edge $y_{r0}^{-}$ of
the - LSL
has exactly the opposite value of that  localized at the edge $y_{r0}^{+}$
of the
+ LSL  where $\Delta y= y_{r0}^{-}-y_{r0}^{+} \gg
\ell_{0}$.  The group velocity of the slow EMP is much smaller than
that of the fast EMP.  Moreover, they can be easily modulated by
applying a transverse electric field to the DQW structure in the case
of strong tunnel coupling.  This renormalization of single-well
fundamental EMPs is essentially different than that of
the right-  and left-edge fundamental EMPs in a single-well 2DEG
channel with relatively short width, $q_{x} W \ll 1$.
As shown in Ref.  \cite{1} , in the latter case the renormalized
EMPs have dispersion laws
$\omega \propto \pm q_{x} \sqrt{\ln(1/q_{x})}$
that give the same absolute value for a group velocity.

We have neglected the possible spatial inhomogeneity of the background
dielectric constant $\epsilon$  along the $z$ direction.  In
GaAlAs/GaAs QWs such an inhomogeneity is relatively small.
Within the RPA  we can include the effect of scattering only in
a phenomenological manner through the small parameter $1/\tau$. This
effect can be treated approximately along the lines of Refs.   \cite{11}
and  \cite{14}. We have also neglected the possible
screening influence of the gates or other free charges outside the
spacer layer. Their influence on the slow   mode, if they   are
at a distance much greater than $\Delta y$ far from
$y_{r0}^{\pm}$,   should be negligible.
We have also assumed that the confining potential, flat in the
interior of the channel, is smooth on the magnetic length
scale but sufficiently steep at the edges that
Landau-level  flattening \cite{21} can be neglected \cite{22}.
In addition, though the thicknesses $d_{l}$ and $d_{r}$ of the QWs have been
neglected, we believe that this often used approximation
  \cite{15} does not affect  our results   essentially.

{\it Note added in proof}. Remarkably the results  of Sec. III A partly
overlap with
the $M=1$ results of Ref. 30, obtained in the time-dependent
Hartree-Fock
 approximation for the particular $M=1$ edge excitations of a double
quantum-dot
system with zero tunneling between the   dots and total   $\nu=2$ in the
limit of large
dots. In this limit the dot radius $R$ is large  compared to both the
magnetic length
$\ell_{0}$ and the dot separation $d$. Then the  energy of the in-phase
collective magnetoplasmon excitations
$E^{+}_{mpl}$ is given  by Eq. (29) of Ref. 30 and that of the
out-of-phase excitations
$E^{-}_{mpl}$ by Eq. (30) of Ref. 30. The former follows from
Eq. (23) and the latter
from Eq. (24) of Sec. III A if we make  the changes $q_{x} \rightarrow
1/R$, $\epsilon \rightarrow 1$ ,
and neglect  $q_{x}V_{g+}$  and the  important terms 0.85 in Eq. (23) and
0.65 in Eq. (24) .
\acknowledgements

We thank Dr. A. H. MacDonald for drawing our attention to Ref. \cite{23}.
This work was supported by the Canadian NSERC Grant No. OGP0121756.  O
G B acknowledges partial support by the Ukrainian SFFI Grant No.
2.4/665.
%\clearpage

\begin{figure}
\caption{Dispersion relation of the DQW fundamental EMPs, for
negligible tunnel coupling $2|T|/\Delta \ll 1$.  The solid curves are for
$ D^{2}/\ell_{0}^{2}=3$ and dashed ones for $D^{2}/\ell_{0}^{2}=10$;
$\omega_{*}=e^{2}/(\pi \hbar \epsilon \ell_{0})$.
The dotted curve is the single-well fundamental
EMP without inter-well Coulomb coupling.  The upper solid and dashed
curves represent the {\it fast } EMP and the lower ones the {\it slow}
EMP.}

\label{fig.1}
\end{figure}
\ \\
\begin{figure}
\caption{Group velocity of the DQW fundamental EMPs for strong tunnel
coupling, $\Delta \leq 2|T|=1$meV, as a function of $\Delta$;
$v_{*}=e^{2}/(\pi \hbar \epsilon)$ and $\Delta_{*}=\sqrt{3}$meV.  The
value of $d/\ell_{0}$ is $0.5$ for the solid and dotted curves and $1$
for the dashed curves.  The upper solid and dashed curves represent
the {\it fast} EMP, for $q_{x} \ell_{0}=10^{-3}$, and the lower ones
the {\it slow} EMP, whose velocity is independent of $q_{x}$.  The
dotted curve represents the {\it fast} EMP for $q_{x} \ell_{0}=10^{-2}$.}

\label{fig.2}
\end{figure}
\ \\
\end{document}